# Fabrication of soft bio-spintronic devices for probing the CISS effect

Ritu Gupta[1], Hariharan V. Chinnasamy[2], Dipak Sahu[3], Saravanan Matheshwaran[2], Chanchal Sow[3] and Prakash Chandra Mondal[1]*

[1]Department of Chemistry, Indian Institute of Technology Kanpur, Uttar Pradesh-208016, India

[2]Department of Biological sciences and Bioengineering, Indian Institute of Technology Kanpur, Uttar Pradesh-208016, India

[3]Department of Physics, Indian Institute of Technology Kanpur, Uttar Pradesh-208016, India

E-mail: pcmondal@iitk.ac.in (P.C.M.)

**Abstract**

Bio-spinterfaces present numerous opportunities to study spintronics across the biomolecules attached to (ferro)magnetic electrodes. While it offers various exciting phenomena to investigate, it's simultaneously challenging to make stable bio-spinterfaces, as biomolecules are sensitive to many factors that it encounters during thin-film growth to device fabrication. The chirality-induced spin-selectivity (CISS) effect is an exciting discovery demonstrating an understanding that a specific electron's spin (either UP or DOWN) passes through a chiral molecule. The present work utilizes Ustilago maydis Rvb2 protein, an ATP-dependent DNA helicase (also known as Reptin) for the fabrication of bio-spintronic devices to investigate spin-selective electron transport through protein. Ferromagnetic materials are well-known for showing spin-polarization, which many chiral and biomolecules can mimic. We report spin-selective electron transmission through Rvb2 that exhibits 30% spin polarization at a low bias (+ 0.5 V) in a device configuration, Ni/Rvb2 protein/ITO measured under two different magnetic configurations. Our findings demonstrate that biomolecules can be put in circuit components without any expensive vacuum deposition for the top contact. Thus, it holds a remarkable potential to advance spin-selective electron transport in other biomolecules such as proteins, and peptides for biomedical applications.

# I. INTRODUCTION

Richard Feynman and Gordon Moore's famous historical statements drove the scientific communities to develop miniaturized electronic devices.[1,2] Wherein molecules are considered the top-level building blocks for device miniaturization.[3–8] Various new analysis techniques and fabrication tools have shifted molecular electronics from science fiction to experimental reality.[9–11] In that, biomolecules such as proteins and peptides have recently gained a lot of interest due to their biochemical and photophysical properties.[12–15] Firstly, it's well-studied fact (by the Ron's group and others) that electron transport through the chiral molecules primarily depend on spin (of electrons) and the chirality of the molecules) both experimentally and theoretically.[16–21] The phenomenon is popularly known as the 'chirality-induced spin selectivity' or CISS effect, suggesting that chiral molecules can behave as efficient spin filters. The CISS effect arises in a chiral molecule due to the coupling of spin and linear momentum of electrons by the chiral electrostatic potential of the biomolecule.[22–25] Thus, the CISS effect encourages us to understand how a preferred electron's spin can transmit through a chiral molecule. The effect has been widely investigated in biomolecules such as DNA, oligopeptides, and proteins in multiheme cytochromes, bacteriorhodopsin, chiral quantum dots, to name a few.[24,26–29] Secondly, since carbon is the most constituent element in biomolecules, thus they have comparatively low spin-orbit coupling (SOC) and hyperfine interaction than the inorganic oxides or metals, thus assists spin-dependent electron propagation in response to an external bias at non-zero voltage. Overall, biomolecules can be the effective spin transport media in solid-state spin-filtering (spintronic) devices due to long spin diffusion length and high spin relaxation time.[12,30,31] Spin-filtering devices utilize specific spin orientation electrons to transport and use selective spin current injection, a typical magnetoresistance devices such as spin-valves.[32–35] With an external magnetic field or ferromagnetic substrate, it is possible to generate enantio-specific substrate-biomolecule interactions and control the transport pathway. However, the previous reports on (bio)spintronic utilizes vacuum deposited metal electrodes on top of the protein films.[36–38] Such a deposition can reduce the device performance by lowering the yield, reproducibility, short-circuiting, and proteins, peptides can get denatured. Thus, we propose here a facile, and versatile top contact that does not require cleanroom facilities but creates a stable contact with the protein. We exploit Ustilago maydis Rvb2 protein to study spin-selective electron transfer in a solid-state device with a configuration of Ni/ Rvb protein/ITO. Rvb1 and Rvb2 are two structurally similar proteins found in eukaryotic cells and belong to the family of AAA+ ATPases (ATPases Associated with diverse cellular activities).[39] They interact with each other to form homo-hexameric or hetero-hexameric structures[40], which can act as chaperones or molecular motors to facilitate various cellular processes.[41] Rvb1 and Rvb2 often function together as components of multiprotein complexes, such as chromatin remodelling complexes.[42] In addition to its role in protein complexes, Rvb2 can also function independently in an antagonistic mode to Rvb1 and regulate transcription.[43] It was shown that Rvb2 interacts genetically with polycomb group (Pcg) gene products and reduces variegation and promotes the spread of heterochromatin[44] in Drosophila. It was reported to inhibit the function of the viral polymerase by interfering with the viral nucleoprotein's oligomerization.[45] Rvb2 also plays a crucial role in T-cell development and regulating humoral mediated immunity.[46] The crystal structure of Rvb2 revealed that it forms hexameric rings, with each subunit adopting a similar conformation.[41] The purified Rvb2 protein was successfully electrostatically attached to a glutathione thin film formed on a ferromagnetic nickel substrate. The spin-specific charge transport is controlled and measured using electrostatically bound Rvb2 protein with glutathione molecules. We demonstrate a case in which the CISS effect not only originates from the chirality of the linker but also from the helical arrangement of the Rvb2 protein, hence providing collective spin polarization. The helical arrangement holds the chiral potential facilitates long-range spin polarization. The work successfully showcases that the Rvb2 protein could be used as an effective spin transport media and provide better spin-selectivity in large solid-state devices and thus could be employed in fabricating biospintronic devices for various other biomolecules.

## II. EXPERIMENTAL

**Materials.** L-Glutathione was purchased from Sigma-Aldrich (Bangalore, India), perchloric acid ($HClO_4$) (Finar), and acetonitrile (Avantore, HPLC-grade) was used as received. Biological 50 mM Tris buffer of pH 8 was prepared in the lab using all chemicals purchased from Sigma-Aldrich (Bangalore, India). For ferromagnetic substrate, nickel (100 nm) was deposited on chromium/Si. Indium tin oxide (ITO, 100 nm) coated glass substrates were purchased from Nanoshel UK Ltd with a sheet resistance of 14–16 Ohm $sq^{-1}$, thickness 1.1 mm and transmittance ∼ 83 % in the visible regime (400-800 nm).

**Protein extraction & purification.**

*U. maydis rvb2* (UMAG_04226) was cloned in pET28a(+) between NdeI and BamHI using the following primers UmRVB2_FP ATGC CATATG GCGCAGATCTCTACCACTTC and UmRvb2_RP ATGC GGATCC TCAAGCTTGAACAGCACCGG. The overexpression of N-terminally His-tagged Rvb2 was carried out in *E.coli* Rosetta grown in L.B. medium. Cultures were grown at 37° C until $OD_{600}$ = 0.6 was reached. It was then induced with 0.8 mM IPTG at 18° C for 12 hours. The protein was purified through three stages of FPLC. The cells from 2 litres of culture were pelleted and resuspended in 100 ml of lysis buffer (50 mM Tris HCl pH 8.0; 150 mM NaCl, 10 mM Imidazole pH 8.0, and 1 mM PMSF). The cells were lysed by probe sonication for 30 minutes (15 s ON and 30 s OFF). The lysate was then centrifuged at 20,000 rpm for 45 minutes at 4 °C, and the supernatant was passed through a 0.22 μm syringe filter. The clarified supernatant was then passed through the Hi Trap Ni-NTA column (Cytiva) and then washed with 20 column volumes (CV) of Buffer W (50 mM Tris HCl pH 8.0; 500 mM NaCl and 20 mM Imidazole pH 8.0). Gradient elution of protein was done with buffers W and E (50 mM Tris HCl pH 8.0; 150 mM NaCl and 500 mM Imidazole pH 8.0). The protein fractions from Ni-NTA were pooled and loaded onto the Hi Trap Heparin column (Cytiva) in buffer A (50 mM Tris HCl pH 7.5; 70 mM NaCl; 2 mM DTT). The heparin column was washed with 20 CV of buffer A, and the protein was gradient eluted with high salt buffer B (50 mM Tris HCl pH 7.5; 1 M NaCl; 2 mM DTT). The eluted protein fractions were concentrated and injected into Superdex 200 10/300 GL (Cytiva) size exclusion chromatographic column equilibrated with buffer S (50 mM Tris HCl pH 7.5; 150 mM NaCl; 2mM DTT).

**Nickel surface modifications:** The nickel substrate was cleaned in hot acetone and ethanol for 15 minutes before surface modification. Afterward, it was treated with 0.1 M $HClO_4$ at $-0.9$ V vs. Ag/AgCl for 20 minutes to reduce the formed oxide layer and further rinsed with nitrogen-purged distilled water to remove unwanted contaminants, similar to a process reported earlier.[47,48] All electrochemical measurement was performed with a conventional three-electrode system using Metrohm Autolab Electrochemical workstation (Model: 204, software nova-2.14). For adsorbing L-Glutathione (GSH) molecules on the nickel surface, 1 µL of GSH, 0.5 mM was at once drop-casted on pre-treated nickel substrate with 0.1 M $HClO_4$ and dried overnight in the desiccator. Later, the GSH-modified Nickel substrate was placed in a glove box with moisture and oxygen levels of less than 5 ppm. After that, the purified Rvb2 protein (30 µM in tris buffer) was electrostatically Attached to the GSH molecules via drop casting onto it and kept for overnight drying. The combined thickness of GSH and Rvb2 protein was determined by Stylus Profiler Model: AlphaStep® D-120 from K.L.A. Tencor. The surface morphology analysis of the substrates Ni/GSH and Ni/GSH/Rvb2 was performed using field emission scanning electron microscopy (FE-SEM) employing Zeiss, model: supra40VP. Static contact angle measurement was performed with Kruss DSA25 goniometer. Approximately 2 µL water droplet was placed for measurement. Contact angle measurement was immediately recorded on both pre-treated nickel and modified nickel. Before and after modifications, magnetic studies of Nickel substrates were studied using custom-made surface magneto-optic Kerr effect.

**Soft-devices fabrication:** A biomolecular spintronic junction was fabricated using Ni/GSH and Ni/GSH-/Rvb2 as bottom electrodes and GEL-coated ITO as the top electrodes. A semi-Gel electrolyte was prepared by mixing 900 mg of $LiClO_4$ and 2.1 g polymethyl methacrylate (PMMA) in 6 mL propylene carbonate and 12 mL acetonitrile with continuous stirring for 3h using vortex.[49] The current-voltage (I-V) studies of all the devices were recorded with Keithley 2604B Source meter.

## III. RESULT AND DISCUSSION

Rvb2 protein is chosen here for spin-dependent electron transport studies due to its helical arrangement, hexameric form, and room temperature stability. The extraction strategy and purification of Rvb2 protein is discussed in the experimental section (Section 1). Purified protein fractions from different chromatographic columns were loaded on 10% SDS-PAGE, and their size exclusion elution profile of Superdex 200 10/300 GL column is also performed. The elution profile conforms to that of a ~350 kDa protein when compared to the elution profile of standards passed through Superdex 200 10/300. The size is indicative of the predominant hexameric form of Rvb2 in solution after size exclusion chromatography. UV-vis absorption spectrum shows clear characteristic peak at 260 and 220 nm due to $\pi$-$\pi^*$ transitions of the electrons in Rvb2 protein. The purified protein was subjected to CD spectroscopy. Near (250-280 nm) and far (195-250 nm) UV spectra of Rvb2 revealed the secondary structure of the protein showing pronounced minima at 208 and 222 nm. The predominance of alpha helices and presence of mixed alpha helices and beta pleated sheets could be gathered from the profile obtained. Ni substrates are highly prone to oxidation and forms a surface oxide layer at ambient conditions, which hampers their ferromagnetic properties.[47] Thus, prior to modifications it was electrochemically treated with 0.1 M $HClO_4$ at − 0.9 V vs. Ag/AgCl for 20 minutes to remove formed oxide layers. Afterward, pre-treated Ni was modified in two steps. Firstly, the pre-treated Ni was modified with GSH molecules via drop-casting and vacuum dried; in the second step, Rvb2 protein was electrostatically assembled and dried in the glove box at $O_2$ concentration <0.5 ppm (**Fig 1a**). The height line profile of Ni/GSH/Rvb2 after scratching, and the average total thickness of Rvb2 from the Ni surface was determined to be 267 ± 12 nm. Later, the structural morphology of the Rvb2 protein was analysed by field emission scanning electron microscopy (FE-SEM) indicating an interconnected China aster flower-like morphology. One can clearly see the hexamer ring form of the Rvb2 protein structure, which correlates well with previous reports for Rvb2 protein structures (**Fig 1b-c**).[39,50,51] The static contact-angle of the water droplet on the Rvb2 protein-modified nickel surface was measured to study the effect of GSH and protein on the hydrophobicity of the nickel surface (Table S1).

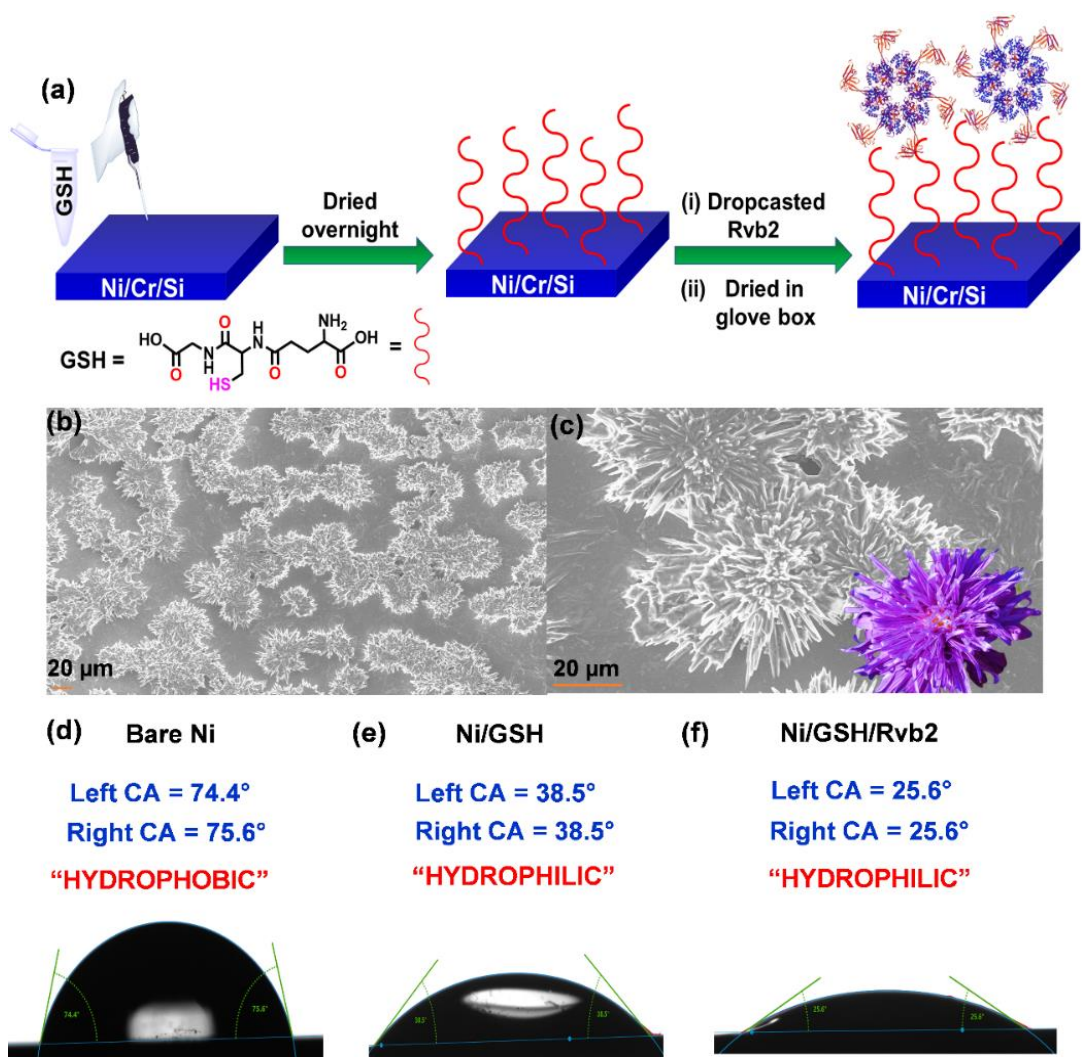

**FIG 1.** (a) Schematic illustration of metallic Ni electrode modification by Glutathione and Rvb2 protein. (b-c) FE-SEM images of Rvb2 electrostatically attracted to Glutathione molecules with high resolution of 500 X (left) and 2 kX (right). Static water contact-angle image of (d) bare Ni, (e) Ni modified with GSH (Ni/GSH), and (f) Ni/GSH modified with Rvb2 protein (Ni/GSH/Rvb2).

Rvb2 protein-modified nickel surface (Ni/GSH/Rvb2) shows hydrophilic surface with a contact angle of 25.6° compared to an electrochemically reduced Ni surface of a contact angle of 75.6° (**Fig 1d,f**). The GSH modified Ni (Ni/GSH) shows a slightly less hydrophilic effect with a contact angle of 38.5° (**Fig 1e**). Rvb2 has more hydrophilic elements such as nitrogen, oxygen, and sulfur than Ni/GSH, hence illustrating a higher hydrophilic character in static contact angle measurement. Further, to understand the magnetization and magneto-optical behaviour, longitudinal MOKE M-H loops were studied at room temperature before and after modification of nickel surfaces. The setup consists of a helium-neon laser source (5mW, 650 nm wavelength), linear polariser (with extinction ratio 9000:1), electromagnet (powered by bipolar power supply), analyser, digital multimeter (DMM), Gaussmeter, silicon photo detector. All the electronic instruments are computer interfaced with LabVIEW programming. The sample is mounted in between the two pole pieces of the electromagnet (as shown) and measurement were performed in a longitudinal mode (external magnetic field parallel to the sample plane). When linearly polarised light is subjected to a magnetised surface (in this case Ni film) the reflected light changes its polarisation and becomes elliptically polarised light.[52] This change in polarisation angle is denoted as Kerr rotation ($\theta_k$) which generally varies from micro-degree to milli-degree. The reflected light is incident on the photo detector. The photo current is measured using a DMM.

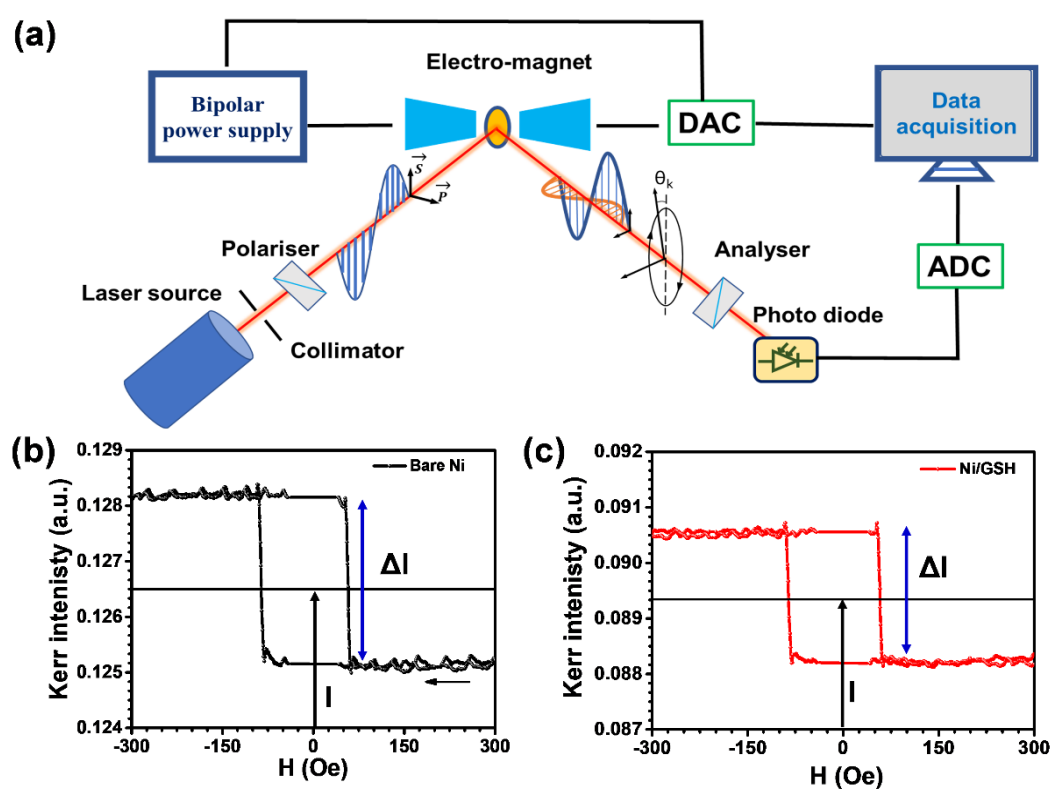

**FIG 2.** (a) Block diagram illustrating MOKE measurement set-up. MOKE hysteresis M-H of (b) e-chem reduced nickel substrate, and (c) Ni/GSH substrate, where I denote mean intensity, and ΔI is the difference between maxima and minima MOKE response.

The measured intensity is a function of $\sin^2(\theta_k)$. While performing experiment without applying magnetic field the polariser and analyser are kept in the crossed state. Application of magnetic field changes the magnetisation direction at the surface which changes the $\theta_k$ accordingly. Thus, $\theta_k$ is proportional to the surface magnetisation. By measuring Kerr intensity/Kerr rotation as a function of magnetic field one can trace the magnetic hysteresis of the surface. When the magnetization switches, the program we have developed allows us to obtain reflected light intensity from the detector (see block diagram in **Fig 2a**). Electrochemically (E-Chem) reduced Ni shows a clear Kerr hysteresis loop with coercivity ($H_c$) of ~57 Oe and average intensity (I) of 0.1264 V (**Fig 2b**). The overall difference in the intensity (ΔI) is ~ 0.003 V. However, due to the presence of a less reflecting non-magnetic GSH layer, Ni/GSH sample shows a drop in the Kerr hysteresis intensity (0.0893 V) while the coercivity ($H_c$) remains unchanged (**Fig 2c**). But the overall difference in ΔI in this case is ~ 0.008 V. This result indicates an enhancement of surface magnetism due to GSH layer. In case of Ni/GSH/Rvb2 performing MOKE experiment was not possible owing to the maximum absorption of light at the surface as non-magnetic Rvb2 surface layer is very thick (266 nm, Fig S7). The spin-specific I-V measurements were accomplished through an external magnetic field generated by placing a permanent magnet of strength (0.3T) at the bottom of the nickel electrode, as shown in **Fig 3a**. The magnetic field was switched "Up" and "Down" to the nickel surface, allowing either parallel or antiparallel velocity vector of the electron to pass through the Rvb2 molecules. The spin polarizability of electrons propagating through the Rvb2 will strongly depend upon the specific spin injected out of Ni and the chirality of the molecule. Also, due to the low spin-orbit interaction of Rvb2, narrow scattering resonance (reference) is experienced by the electrons travelling through the Rvb2 protein. Spin-dependent electrical measurements (I-V & semilog I-V) data for Rvb2 biomolecular junction with and without magnetic field are shown in **Fig 3b-c**. In both cases, the junction shows an asymmetric I-V behaviour, possibly due to the asymmetric nature of the two electrodes (Ni as the bottom and ITO as the top electrode). Interestingly, a clear increment in current density for the "UP" magnet direction compared to the "DOWN" magnet direction is observed in the I-V response of biomolecular junction, a clear signature of the CISS effect. Also, the current density for the "No" magnet condition was comparatively similar to that for the "Down" magnet. The threshold for the current at non-zero bias for two different spins can be obtained from the semilog plot (shown in the inset, **Fig 3c**). The control experiment without GSH and Rvb2 protein was also studied, and an evident linear ohmic behaviour was observed. Similarly, I-V measurement with GSH molecules sandwiched between Ni and gel-ITO was also conducted, and a slight non-ohmic behavior was obtained. Further, to quantify asymmetry in the current data plot for Ni/GSH/Rvb2/Gel/ITO biomolecular junction, spin polarization (SP) was calculated by using the following well-known equation 1, where $I(V)_\uparrow$ is current at a particular voltage when the magnetic field is applied 'UP' direction and $I(V)_\downarrow$ is when the magnetic field is in the 'DOWN' direction.[53–55]

$$SP(V) = \frac{I(V)_\uparrow - I(V)_\downarrow}{I(V)_\uparrow + I(V)_\downarrow} \quad (1)$$

The behaviour of spin polarization as a function of applied voltage for the Rvb2 junction is shown in **Fig 3d**. The Rvb2 protein shows a minimum SP of 10% at a less threshold voltage of 100 mV, whereas a maximum SP of 30% was observed at 500 mV. Thus, the CISS effect is highly bias-dependence, in addition to the chirality of the system.

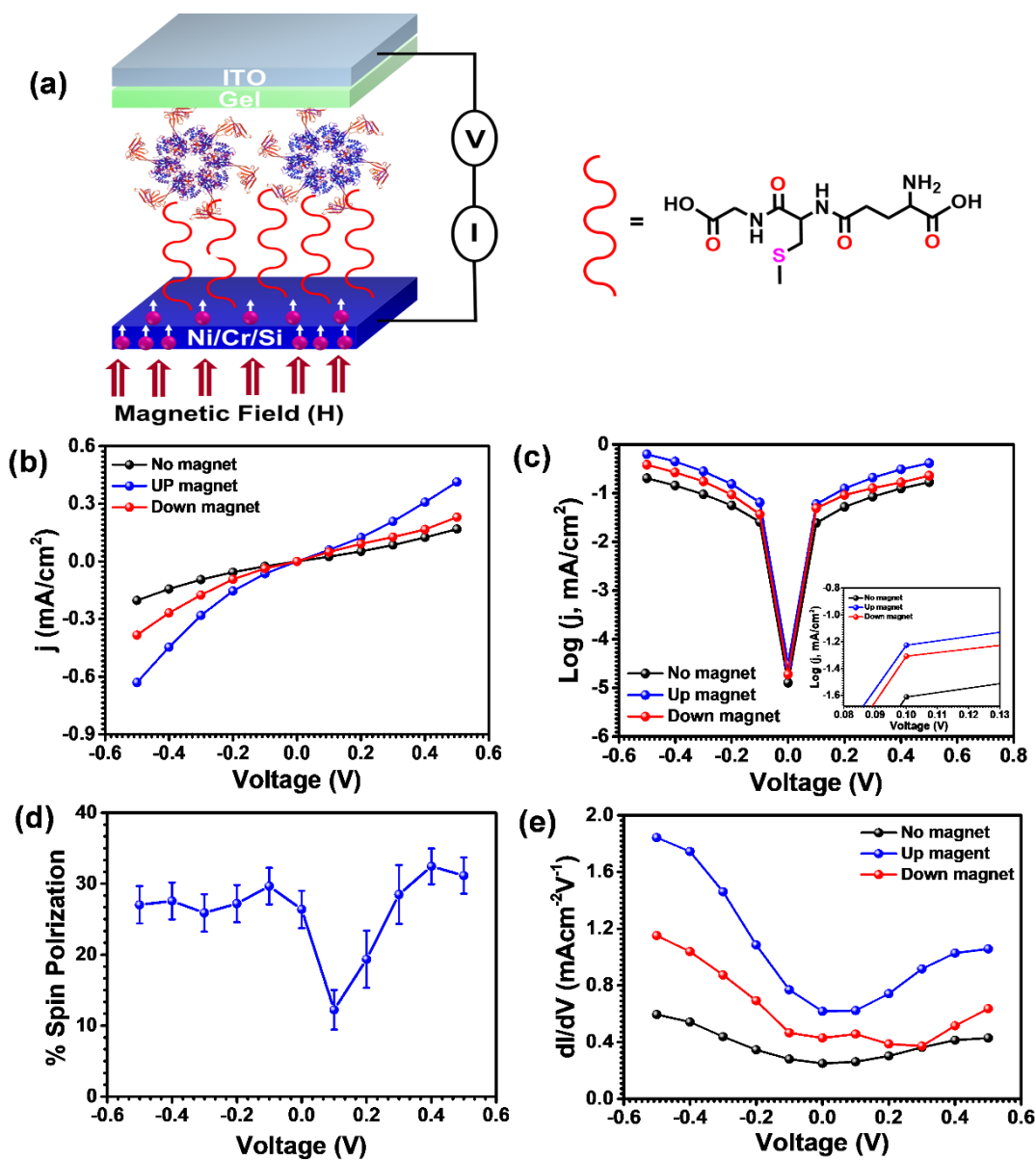

**FIG 3.** (a) Schematic image of the Ni/GSH/Rvb2/Gel/ITO. biomolecular junction. (b) I-V characteristics of the junction in the presence of with and without magnetic field. (C) Semilog plot of the corresponding junction. (d) % Spin polarization at different voltage with error bars. (e) differential conductance (dI/dV) vs V plot of the junction for bias voltage − 0.5 to 0.5 V.

Apart from that, the differential conductance (dI/dV) vs. V profile is constructed to illustrate the local densities of states (LDOS) in the "No" magnet condition and "UP" and "DOWN" magnet conditions (**Fig 3e**). The (dI/dV) vs. V reveals the asymmetric nature of the junction for both conditions, as the LDOS are higher in negative bias than the positive bias. In the case of the "DOWN" magnet, the LDOS remains constant from 0 V up to ~ 200 mV in positive bias and then regularly increases up to 500 mV.

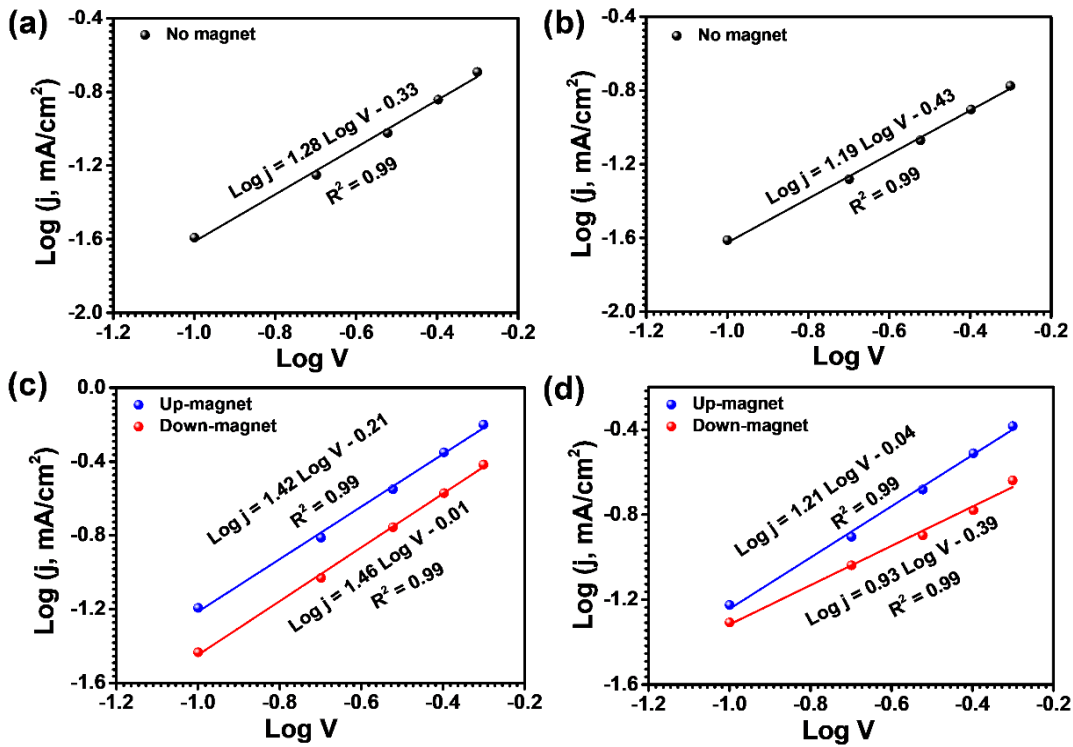

**FIG 4.** Log j vs. Log V plot of the bio-molecular junction without magnet (a) in negative bias and (b) positive bias range. Log j vs Log V plot of the junction with a magnet (c) in negative and (d) positive bias range.

At the same time, LDOS continuously increases in negative bias, hence showing higher conductance at negative biasing. Similar behaviour was observed for the "No" magnet condition with a slight decrease in conductance. However, when the magnet was kept in the "UP" direction, the LDOS was much higher than the "DOWN" condition and frequently increased in both positive and negative bias leading to higher conductance at the UP-magnet direction compared to the DOWN magnet condition. This type of behavior for chiral molecules is well-known both experimentally and theoretically.[21,56] Further, to understand the selective charge transport mechanism Log j vs. Log V plot was extracted from corresponding I-V plots without and with a magnet. In the absence of an external magnetic field, the junction shows a linear dependence on voltage in both negative and positive bias range up to 0.5 V following the direct tunnelling under Simmon's approximations (**Fig 4a-b**). However, in the "UP" magnet condition, the junction exhibits that current depends on the bias with a relationship, $I = V^{3/2}$ in both the bias range following a classical child-Langmuir law respectively (**Fig 4c**). Contrarily, in the case of the "DOWN" magnet condition, again, a linear dependence on voltage was observed following the similar Simmon's approximations at both the bias range (**Fig 4d**).[57] With all this experimental evidence, we proposed a mechanistic pathway behind this spin-selective charge transport in Rvb2 protein. Being-ferromagnet, nickel has different DOS (density of states) at the Fermi level for spin-UP and spin-DOWN electrons without an external magnetic field, as shown in **Fig 5a**. In the case of the UP-magnet condition, the DOS of spin-UP electrons increases, allowing only spin-UP electrons to pass through the GSH/Rvb2 layers. Also, the spin-filter nature of proteins due to the CISS effect selectively permits the spin-UP electron to transport through the Rvb2 and reach the top contact (**Fig 5b**). On another side, very few spin-DOWN electrons transport through the layers, thus providing lesser current than the UP-magnet condition (**Fig 5c**).

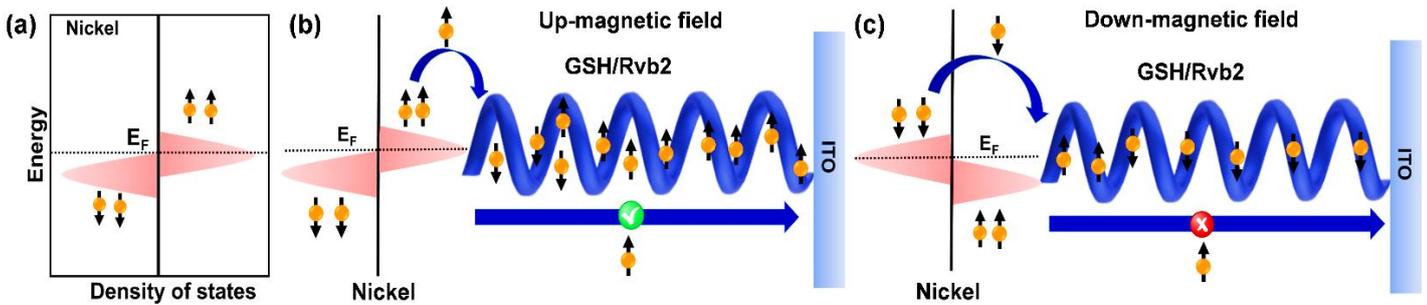

**FIG 5.** (a) E vs. DOS for a ferromagnetic material in the absence of an external magnetic field, (b) Proposed mechanism on the spin-selectivity in the protein in two different magnetic conditions, spin UP, and DOWN.

## IV. CONCLUSION

We have demonstrated the construction of the Rvb2-based biomolecular junction, soft-spintronic devices without any need for expensive vacuum deposition for the top electrical contact. The hexametric ring, and helical arrangement of the Rvb2 on the Ni electrode surface makes the protein suitable for the CISS effect investigated at the room temperatuere. A covalent bond at the ferromagnet/chiral molecules interfaces was made via a chemical approach and this is crucial, that would help spin diffusion longer and without spin decoherent. Besides, protein-induced surface polarization helps to have efficient spin-propagation. Higher current density and differential conductance were obtained by switching the magnet from the "DOWN" to "UP" direction, as compared to the no protein, a clear demonstration of the CISS effect that yields high spin polarization of 30% at 0.5 V at room temperature even though a relatively thicker protein (~ 266 ± 12 nm) thickness layer was incorporated in between two electrical contacts. The long-range spin-selectivity is observed to be related to the preferred handedness of the GSH, Rvb2 protein, and less spin-scattering. The work provides new insights into the spin effect on the Rvb2 protein. The obtained long-range spin selectivity enables the room-temperature operation of spintronic devices, an essential requirement for practical applications. Though many studies, including theory and experiment, describe the CISS effect, still, it remains a puzzling problem. Thus, more theoretical frameworks and experimental tools are desirable for a better understanding of spin-dependent charge conduction mechanisms to realize the CISS phenomena in-depth.

## V. ACKNOWLEDGMENTS

R.G. thanks I.I.T. Kanpur for a Ph.D. fellowship. P.C.M. acknowledges financial support from Science and Engineering Research Board (Grant No. C.R.G./2022/005325), New Delhi, India. The authors acknowledge I.I.T. Kanpur for infrastructures and equipment facilities. A great help from Dr. Ankur Malik for the TOC artwork, and Ranjeev Parashar for providing gel electrolyte is highly appreciated. Authors thank Prof. Ron Naaman and Ms. Kakali Santra (WIS, Israel) for providing Ni substrates.

## AUTHOR DECLARATIONS

Conflict of Interest

The authors have no conflicts to disclose.

## DATA AVAILABILITY

Data available on request from the corresponding author.